\documentclass{aa}
\usepackage{psfig}

\begin{document}

   \thesaurus{06         % A&A Section 6: Form. struct. and evolut. of stars
              (19.53.1;  % Stars: oscillations of,
               19.63.1)} % Stars: structure of.
   \title{The distance modulus of the Small Magellanic Cloud based on 
          double-mode Cepheids}

   \subtitle{}

   \author{G. Kov\'acs
          \inst{}
%          \and
%          C. Ptolemy\inst{2}\fnmsep\thanks{Just to show the usage
%          of the elements in the author field}
          }

   \offprints{G. Kov\'acs}

   \institute{Konkoly Observatory, P.O. Box 67, H-1525, Budapest, Hungary \\
              email: kovacs@konkoly.hu
%         \and
%             University of Alexandria, Department of Geography\\
%             email: c.ptolemy@hipparch.uheaven.space
%             \thanks{The university of heaven temporarily does not
%                     accept e-mails}
             }

   \date{Received June 2, 2000; accepted July 6, 2000}

   \maketitle

   \begin{abstract}

We employ the very recent photometric data of the {\sc ogle} project 
together with stellar atmosphere and linear pulsation models to 
determine the distance modulus of the Small Magellanic Cloud  
from its double-mode Cepheids. Based on the requirement of obtaining 
the same distance modulus $DM$ from the two types of variables 
(fundamental \& first overtone and first \& second overtone), we get 
$DM=19.05$~mag, with a very small statistical error (standard 
deviation) of $0.017$~mag. Various systematic and zero point 
ambiguities (primarily those of the color--temperature 
transformation) lead to an error of $\pm 0.13$~mag (estimated $3\sigma$  
deviation). This result is in very good agreement with the distance 
modulus of the Large Magellanic Cloud of $18.5$~mag, derived earlier 
from cluster double-mode RR~Lyrae stars. 

   \end{abstract}

%
%________________________________________________________________

\section{Introduction}

In a former paper (Kov\'acs \& Walker 1999, hereafter KW99) we have 
shown that the distances derived from double-mode RR~Lyrae (RRd) stars 
are systematically larger than the ones obtained from the standard 
Baade-Wesselink (BW) analyses of RR~Lyrae stars. The latter 
distances are in similar conflict also with the BW results of 
Cepheids (e.g., Gieren et al.~1998). The reason of 
this discrepancy is still unknown (Cacciari et al.~2000). This 
contributes further to the ambiguity at the $0.2$--$0.3$~mag level 
over the luminosity scale of the RR~Lyrae stars, and consequently, 
over the distances to the nearest globular clusters and galaxies. 

The purpose of this Letter is to study the applicability of double-mode 
variables in the distance calibration in more detail. The discovery of 
a large sample of fundamental \& first overtone (FU/FO) and first 
\& second overtone (FO/SO) Cepheids in the Small Magellanic Cloud 
(SMC) by the {\sc ogle} team (Udalski et al. 1999a, hereafter U99) 
enables us to derive the distance to the Cloud and thereby adding 
another piece of information to the dispute over the distance of 
the Magellanic Clouds. 

%__________________________________________________________________

\section{Data, method, models and temperature scales}

Double-mode (or beat) Cepheids in the SMC have been discovered 
previously by the major microlensing projects ({\sc macho}, 
Alcock et al.~1997; {\sc eros}, Beaulieu et al.~1997). However, 
it is only the {\sc ogle} team who publishes the data in 
standard colors (i.e., in Johnson $V$ and in Kron-Cousins $I_c$). 
Therefore, for the time being, we decided to employ their data only. 

We use the periods, average $V-I_c$ colors and $V$ magnitudes of the 
23 FU/FO and 70 FO/SO Cepheids published by 
U99.\footnote {Actually, we use the slightly revised data set, see 
the {\sc ogle} {\it ftp} site at {\it sirius.astrouw.edu.pl}} 
Although they presented also $B-V$ colors, we decided not to use them, 
because of the few data points in $B$. For reddening we accepted their 
values derived for the various fields from a method based on red clump 
stars (see U99 and the {\it ftp} site mentioned above).    

In computing the distance modulus, we follow almost entirely the 
method of KW99. Here we repeat only the basic steps and assumptions. 

For any double-mode variable, from the pulsation models we obtain 
relations between the physical parameters and the periods 
%
%%%%%%%%%%%%%%%%%
%    Eq. (1)  
%%%%%%%%%%%%%%%%%
%
\begin{eqnarray}
P_i = f_i(M,L,T_{\rm eff},X,Z) \hskip 2mm ,
\end{eqnarray}
where $i=0,1$ or $1,2$ for the FU/FO and FO/SO variables, respectively. 
The other parameters have their usual meaning. In principle, from the 
observed color we can determine $T_{\rm eff}$ and from other information 
we also have approximate values for the hydrogen and metal abundances. 
Therefore, we can invert the above relations to compute the 
luminosity (and mass). Next, the distance modulus $DM$ is computed 
simply from the comparison of the observed $V$ magnitude and the 
calculated absolute magnitude from $L$ by using a bolometric 
correction ($B.C.$) formula and proper interstellar reddening. The 
existence of the two types of beat Cepheids in SMC enables us to 
estimate $DM$ in two different ways. Because of the different 
dependence of the periods on the physical parameters in the two types 
of variables, this will allow us to calculate $DM$ in an optimum way 
and put constraints on $Z$, one of the most important parameters entering 
in this method. The other crucial parameter is the {\it zero point} 
of $T_{\rm eff}$, which cannot be constrained from the present data, but, 
in principle, it is also possible to do. 

The main assumptions entering in our approach are the following:

\begin{itemize}
\item{}
FU/FO and FO/SO variables have the same chemical composition.
\item{}
Linear nonadiabatic and purely radiative pulsation model periods are 
applicable to the observed periods.
\item{}
The SMC has small spatial extent relative to its distance.
\item{}
Current color--temperature calibrations are reliable enough.
\end{itemize} 

As regards the technical details, it should be mentioned that we used 
the same standard, purely radiative linear pulsation code as in KW99. 
A large number of models were computed with $X=0.76$ and $Z=0.001$, 
$0.002$, $0.003$, $0.004$, $0.008$ and using {\sc opal'96} opacities 
(Iglesias \& Rogers 1996). The distribution of the various elements 
in $Z$ corresponds to that of the Sun. Unlike in the case of RRd models, 
we could not find a simple way to invert Eq.~(1). Therefore, in finding 
the values of $(M,L)$ which fit the observed periods the best for the 
given $T_{\rm eff}$, $X$ and $Z$, we used a straightforward search 
in the interpolated fine grid of the original model sequences. 

As far as the $(V-I_c)\rightarrow T_{\rm eff}$ and $B.C.$ calibrations are 
concerned, we used the stellar atmosphere models of Castelli et al.~(1997) 
and obtained the following formulae through least squares fits 
in a parameter space relevant for beat Cepheids (i.e., 
$T_{\rm eff}=5000-7000$K, $\log~g=2.0-3.5$, $\rm{[M/H]}=-1.5-0.0$)
%
%%%%%%%%%%%%%%%%%%%%%
%    Eqs. (2) & (3)  
%%%%%%%%%%%%%%%%%%%%%
%
\begin{eqnarray}
\log T_{\rm eff} & = & 3.9224 - 0.2470(V-I_c) + 0.0046\log g \nonumber \\  
             & + & 0.0012{\rm [M/H]} \hskip 2mm , \\
B.C.         & = & 0.0411 + 2.0727\Delta T - 0.0274\log g \nonumber \\ 
             & + & 0.0482{\rm [M/H]} - 8.0634\Delta T^2 \hskip 2mm ,
\end{eqnarray}
where $\Delta T=\log T_{\rm eff}-3.7720$, and we used the standard notation 
for the relative heavy element abundance [M/H]$=\log Z/Z_{\odot}$, with 
$Z_{\odot}=0.02$. 

The above $B.C.$ is adjusted to $M_{\rm bol}(\odot) = 4.75$ and yields 
$B.C.(\odot)=-0.09$. The zero point of Eq.~(2) is only marginally 
higher than that of Blackwell \& Lynas-Gray (1994, hereafter BLG94) 
(see also Clementini et al.~1995), which is based on the InfraRed 
Flux Method (IRFM). Other, more recent IRFM-based $T_{\rm eff}$ 
calibrations (Blackwell \& Lynas-Gray 1998, hereafter BLG98; 
Alonso et al. 1999, hereafter A99) yield lower zero points 
($\log T_{\rm eff} (BLG98) - \log T_{\rm eff} (BLG94) \approx -0.004$, 
$\log T_{\rm eff} (A99) - \log T_{\rm eff} (BLG94) \approx -0.010$). We caution 
however, that these differences are based on  $(V-I_c)\rightarrow T_{\rm eff}$  
calibrations, and the corresponding formula of A99 can be employed 
only after applying a transformation from their Johnson $I$ to $I_c$ 
(Fernie 1983). This could perhaps be one of the reasons why we get 
(somewhat curiously) better agreement between BLG94 and A99 with 
the temperatures calibrated by $B-V$. In the next section we will 
check the sensitivity of the derived distance modulus against the 
various $T_{\rm eff}$ scales. 

Finally we note that for the calculation of the gravity  we used the 
following formula
%
%%%%%%%%%%%%%%%%%%%%%
%    Eq. (4) 
%%%%%%%%%%%%%%%%%%%%%
%
\begin{eqnarray}
\log g & = & 2.62 - 1.21\log P_{FU} \hskip 2mm ,
\end{eqnarray}
which has been derived from a simple pulsation equation and black-body 
relation. This formula has a maximum error of $\pm 0.05$, assuming 
that $M/M_{\odot}=3.0^{+2.0}_{-1.0}$.

\section{Determination of the distance modulus}

To characterize the quality of the fit, for each variable, we calculated 
the following quantity  
%
%%%%%%%%%%%%%%%%%%%%%
%    Eq. (5) 
%%%%%%%%%%%%%%%%%%%%%
%
\begin{eqnarray}
\sigma = \sqrt{\Delta P_i^2 + \Delta P_{i+1}^2} \hskip 2mm ,
\end{eqnarray}
where $i=0$ or $1$ for the FU/FO and FO/SO variables, respectively. 
The difference between the observed and calculated periods are denoted 
by $\Delta P_i$. We found basic difference between the two types of 
variables in respect of the behavior of $\sigma(M,L)$. In Figs.~1 and 2 
we show representative gray maps and one dimensional slices along the 
minimum values of $\sigma$. For better visibility, in the gray maps we 
used a relatively low resolution, and therefore we considered only the 
minimum value of $\sigma$ in each pixel (the scans were performed on a 
much finer grid). We see that the minimum (indicated by the black pixels) 
is much more pronounced for the FU/FO than for the FO/SO variables. 
This is nicely seen also in the slices along the minimum values. 
In addition, there is also a slight offset of $<0.002$ for $P_2/P_1$ 
between the observed and theoretical values. Although this problem 
might bear some theoretical significance, it is completely unimportant 
in the present context (see Table 1). Furthermore, the shallow minima of 
the FO/SO variables yielded more stable $DM$s, depending (much) less on 
the various parameter changes than those of the FU/FO variables. This, 
together with their large number, make them very valuable for the purpose 
of distance estimation. 

%
%>>>>>>>>>>>>>>>>>>>>
%
%   FIGURE 1.
%
%>>>>>>>>>>>>>>>>>>>>
%
\begin{figure}[t]
\vskip -0mm
\centerline{\psfig{figure=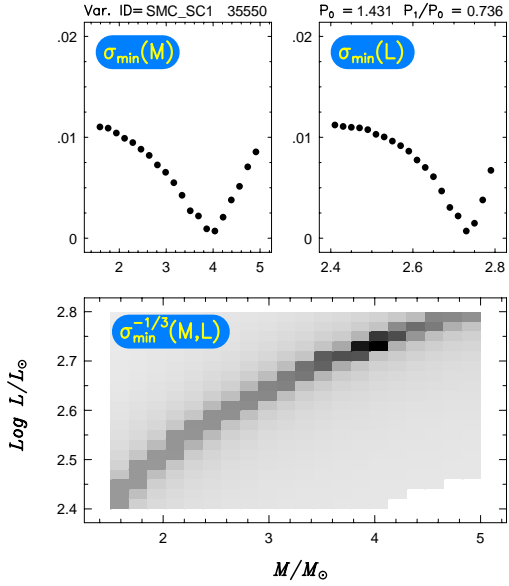,height=80mm,width=70mm}}
\vskip -0mm
\caption{
{\it Lower panel:} Gray map of the period deviation (Eq.~(5)) in 
the $(M,L)$ parameter space. Darker areas indicate better fits to 
the observed periods. 
{\it Upper panels:} minimum period deviations as functions of 
$M$ and $\log L$. An FU/FO variable is tested.   
}
\vskip 0mm
\end{figure}
%

%
%>>>>>>>>>>>>>>>>>>>>
%
%   FIGURE 2.
%
%>>>>>>>>>>>>>>>>>>>>
%
\begin{figure}[t]
\vskip -0mm
\centerline{\psfig{figure=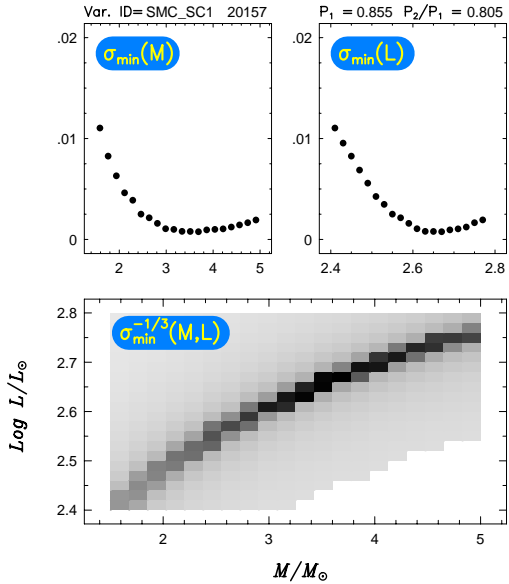,height=80mm,width=70mm}}
\vskip -0mm
\caption{
As in Fig.~1, but for an FO/SO variable.   
}
\vskip 0mm
\end{figure}

The calculation of the distance modulus was performed for each $Z$ 
listed in the previous section. By applying the temperature scale 
given by Eq.~(2), the individual $DM$s are plotted in 
Figs.~3 and 4. We employed the $3\sigma$ criterion for filtering 
out a few outliers (the number of these variables never exceeded 
two). The most striking feature of these plots is the large 
sensitivity of $DM_{01}$ ($DM$ of the FU/FO variables) against the 
variation of $Z$. This is in contrast with the very {\it weak 
sensitivity} of $DM_{12}$. These properties and the $DM_{01}=DM_{12}$ 
condition allow us an optimum estimation of $Z$. 
(We note in passing that the other natural criterion, minimum 
dispersion of all individual distance moduli, leads to the same 
conclusion.) As it is seen in Fig.~4, the present results prefer a 
somewhat lower abundance than the one most often used in the context 
of the SMC Cepheids. With $Z_{\odot}=0.02$, our close to optimum $Z$ 
corresponds to [Fe/H]$=-0.8$, whereas the usually quoted value is $-0.7$ 
(corresponding to $Z=0.004$, see Luck et al. 1998). We think that 
this is a fair agreement and a good sign of the consistency between 
the present, completely independent estimation of $Z$ and those 
obtained by direct spectroscopic observations (of other Cepheid 
variables). 

%
%>>>>>>>>>>>>>>>>>>>>
%
%   FIGURE 3.
%
%>>>>>>>>>>>>>>>>>>>>
%
\begin{figure}[t]
\vskip -0mm
\centerline{\psfig{figure=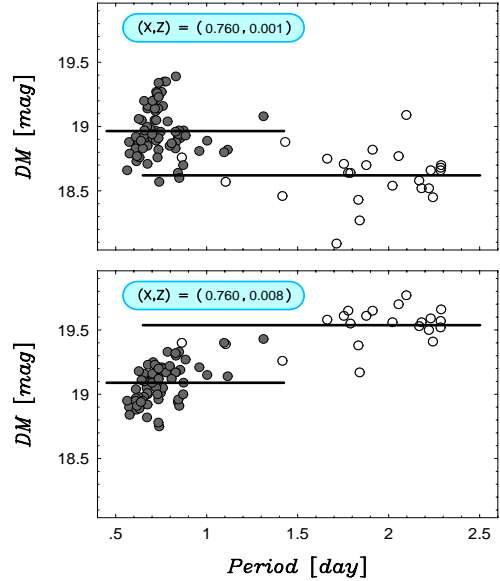,height=85mm,width=70mm}}
\vskip -5mm
\caption{
Individual and average distance moduli calculated for the FU/FO 
and FO/SO variables (shown by open and gray filled circles, respectively). 
Average $DM$s are shown by thick lines. Chemical compositions of the 
models used are given in the top left corners.    
}
\vskip 0mm
\end{figure}
%

%
%>>>>>>>>>>>>>>>>>>>>
%
%   FIGURE 4.
%
%>>>>>>>>>>>>>>>>>>>>
%
\begin{figure}[t]
\vskip -0mm
\centerline{\psfig{figure=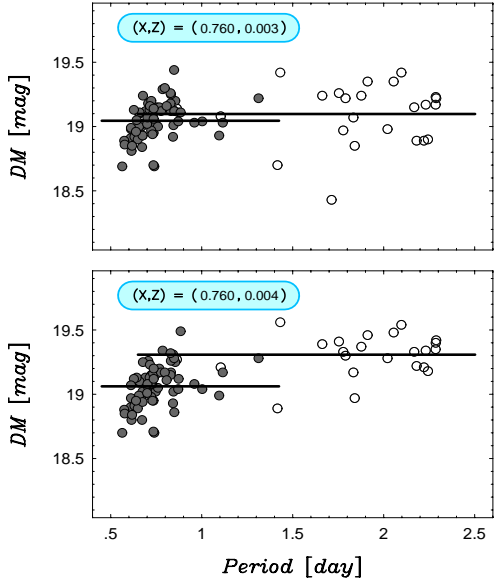,height=85mm,width=70mm}}
\vskip -5mm
\caption{
As in Fig.~3, but for different heavy element abundances.  
}
\vskip 0mm
\end{figure}

Although Fig.~4 shows that the exact $DM_{01}=DM_{12}$ condition 
is still not satisfied for $Z=0.003$, and a somewhat lower $Z$ 
would be more appropriate, the difference is within the reasonable 
error limit, and therefore, in the estimation of $DM$, we use 
the result obtained at $Z=0.003$. When weighted by the number of 
stars, we get $19.05$~mag for the average rounded distance modulus. 
By using the lower $T_{\rm eff}$ scale of A99, a similar match is 
found at $Z=0.003$ between $DM_{01}$ and $DM_{12}$. In this case 
we get $DM=18.90$~mag. Due to the large number of variables, the 
formal statistical errors are very small in both cases: 
$\sigma_{\rm DM}=0.017$~mag. Much more significant sources of errors are 
the various systematic effects and zero point ambiguities. Table 1 
summarizes the changes in the distance moduli caused by the most 
significant potential sources of these kinds of errors. 
%
%>>>>>>>>>>>>>>>>>>>>
%
%   TABLE 1.
%
%>>>>>>>>>>>>>>>>>>>>
%
\begin{table}[ht]
\caption{Systematic errors in the distance moduli at $Z~=~0.003$}
\begin{flushleft}
\begin{tabular}{llll}
\hline
Quantity & & $\Delta DM_{01}$ & $\Delta DM_{12}$ \cr   
\hline
$\Delta \log T_{\rm eff}$   & $ = +0.01  $  & $+0.22$ & $+0.11$ \cr
$\Delta V           $   & $ = +0.02  $  & $-0.06$ & $-0.03$ \cr
$\Delta I_c         $   & $ = +0.02  $  & $+0.10$ & $+0.05$ \cr
$\Delta E_{B-V}     $   & $ = +0.05  $  & $+0.15$ & $+0.02$ \cr
$\Delta P_0         $   & $ = +0.01  $  & $-0.00$ & $.........$ \cr
$\Delta P_1/P_0     $   & $ = +0.002 $  & $-0.15$ & $.........$ \cr
$\Delta P_1         $   & $ = +0.01  $  & $.........$ & $-0.04$ \cr
$\Delta P_2/P_1     $   & $ = +0.002 $  & $.........$ & $-0.01$ \cr
\hline
\end{tabular}
\end{flushleft}
{\footnotesize
\underline{Note:}
The period errors refer to the model values.}
\end{table}
It is difficult to assess the size of the systematic errors in the 
various quantities. The numbers entering in the table are our best 
guesses on the $3\sigma$ errors (for notation simplicity we used 
positive changes everywhere). The assumed ambiguity in $T_{\rm eff}$ is 
based on the difference between A99 and BLG94. We think that this is 
a generous overestimation of the true error, because of the 
ambiguities mentioned in Sect.~2 and because of the smaller difference 
obtained in a comparison with the other current scale of BLG98 
(see Sect.~2). 

The following conclusions can be drawn from the table: 

\begin{itemize}
\item[(a)]
FU/FO variables are (much) more sensitive to the systematic changes 
than FO/SO variables.  
\item[(b)]
The most serious source of error is the zero point ambiguity in the 
$T_{\rm eff}$ scale.
\item[(c)]
Considering only the FO/SO variables, which dominate the value of the 
average distance modulus, and assuming that the various errors are 
independent, we get $\pm0.13$~mag for the total estimated systematic 
error. 
\end{itemize}

\section{Conclusions}

By using the relative distance of $0.51$~mag determined by Udalski 
et al.~(1999b), the present determination of the SMC distance leads to an 
LMC distance modulus of $18.54$~mag. This value is magically close to 
the value of $18.53$~mag, derived from the application of the same method 
to Galactic double-mode RR~Lyrae stars and using the relative distance 
of a few LMC globular clusters (Kov\'acs 2000). Considering that this 
result was derived on a completely different data set, we think that 
the agreement is remarkable. This result suggest that the present models 
and input physics are more compatible with the observations than implied 
by Buchler et al.~(1996) from their beat and bump Cepheid studies. 

As given in Table 1, the most important source of ambiguity in this method 
is the potential error in the zero point of the temperature scale. Even if 
we consider this ambiguity, it is not possible to lower the distance modulus 
by more than $\approx 0.13$~mag. This emphasizes further the contradiction 
between this `long' and other `short' distances, obtained e.g., by 
statistical parallax and red clump methods (Udalski et al.~1999b, see 
however Romaniello et al.~2000).

\begin{acknowledgements}
We are grateful to Andrzej Udalski for additional information 
about the {\sc ogle} data quality. Production of the corresponding 
interpolated opacity tables by Zolt\'an Koll\'ath is thanked. 
Fruitful discussions with Robert Buchler are appreciated.    
The supports of the following grants are acknowledged: 
{\sc otka t$-$024022, t$-$026031} and {\sc t$-$030954}.
\end{acknowledgements}


\begin{thebibliography}{}

\bibitem{} Alcock C. \& the {\sc macho} collaboration, 1997, 
           astro-ph/9709025 
\bibitem{} Alonso A., Arribas S., Mart\'\i nez-Roger C., 1999, 
           A\&AS 140, 261 (A99)
\bibitem{} Beaulieu J. P. \& the {\sc eros} collaboration, 1997, A\&A 321, L5
\bibitem{} Blackwell D. E., Lynas-Gray A. E., 1994, A\&A 282, 899 (BLG94)
\bibitem{} Blackwell D. E., Lynas-Gray A. E., 1998, A\&AS 129, 505 (BLG98)
\bibitem{} Buchler J. R., Koll\'ath Z., Beaulieu J. P., Goupil M-J., 1996, 
           ApJ 462, L83
\bibitem{} Cacciari C., Clementini G., Castelli F., Melandri F.,  
           2000, ASP Conf. Ser. 203, 176 
\bibitem{} Castelli F., Gratton R. G., Kurucz R. L., 1997, A\&A 318, 841
\bibitem{} Clementini G., Carretta E., Gratton R., Merighi R., 
           Mould J. R., McCarthy J. K. 1995, AJ 110, 2319
\bibitem{} Fernie J. D., 1983, PASP 95, 782
\bibitem{} Gieren W. P., Fouqu\'e P., G\'omez M., 1998, ApJ 496, 17
\bibitem{} Iglesias C. A., Rogers F. J., 1996, ApJ 464, 943 
\bibitem{} Kov\'acs G., Walker A. R., 1999, ApJ 512, 271 (KW99)
\bibitem{} Kov\'acs G., 2000, to appear in ASSL Series, Eds. M. Takeuti, 
           D. D. Sasselov
\bibitem{} Luck R. E., Moffett T. J., Barnes T. G., Gieren W. P. 1998, 
           AJ 115, 605 
\bibitem{} Romaniello M., Salaris M., Cassisi S., Panagia N., 2000,
           ApJ 530, 738
\bibitem{} Udalski A., Soszy\'nski I., Szyma\'nski M., Kubiak M., 
           Pietrzy\'nski G., Wo\'zniak P., \.Zebru\'n K., 1999a, 
           Acta Astron. 49, 1
\bibitem{} Udalski A., Szyma\'nski M., Kubiak M., Pietrzy\'nski G., 
           Soszy\'nski I., Wo\'zniak P., \.Zebru\'n K., 1999b, 
           Acta Astron. 49, 201

\end{thebibliography}
\end{document}